\documentstyle[preprint,prl,tighten,aps,epsfig]{revtex}    %

\begin{document}

\draft

\preprint{\begin{tabular}{l}
\hbox to\hsize{\mbox{ }\hfill DESY 99-046}\\
\hbox to\hsize{\mbox{ }\hfill KIAS-P99021}\\
\hbox to\hsize{\mbox{ }\hfill SNUTP 99-019}\\
\hbox to\hsize{\mbox{ }\hfill hep--ph/9904276}\\
\hbox to\hsize{\mbox{ }\hfill June, 1999}\\
          \end{tabular} }

\title{CP Phases in Supersymmetric Tri--lepton Signals at the Tevatron}

\author{S.Y.~Choi}
\address{Korea Institute for Advanced Study, 207--43, Cheongryangri--dong
         Dongdaemun--gu, Seoul 130--012, Korea}

\author{M.~Guchait} 
\address{Deutsches Elecktronen--Synchrotron (DESY), D--22603 Hamburg, Germany }

\author{H.S.~Song and W.Y.~Song}
\address{Center for Theoretical Physics and Department of Physics
         Seoul National University, Seoul 151-742, Korea}

\maketitle

\begin{abstract}
We have analyzed the supersymmetric tri--lepton signals for sparticle 
searches at the Tevatron in the minimal supersymmetric standard model  
with general CP phases without generational mixing. The CP phases may 
affect very strongly the chargino and neutralino mass spectrums
and $\sigma(p\bar{p}\rightarrow\tilde{\chi}^-_1\tilde{\chi}^0_2)$ as well 
as ${\cal B}(\tilde{\chi}^-_1\rightarrow\tilde{\chi}^0_1\ell^-\nu)$
and ${\cal B}(\tilde{\chi}^0_2\rightarrow\tilde{\chi}^0_1\ell^+\ell^-)$.
Even under the stringent constraints from the electron electric dipole moment
the CP phases can lead to a minimum of the tri--lepton event rate 
for their non-trivial values.
\end{abstract}



The low--energy minimal supersymmetric standard model (MSSM) \cite{MSSM} 
in general involves a large number of CP--violating phases. 
Nevertheless, their presence has largely been ignored in phenomenological 
analyses because of the complexity due to the introduction of many 
new independent parameters and because of severe constraints on 
individual phases by the experimental limits for electron and neutron 
electric dipole moments (EDM) obtained by fixing other phases to zero 
\cite{MS}. However, many recent 
works \cite{KO,Kaplan,IN} have shown that these constraints could be evaded 
without suppressing the CP--violating phases of the theory. 
One option is to make the first two generations of scalar fermions
rather heavy so that one--loop EDM constraints are automatically evaded. 
As a matter of fact one can consider so--called effective
SUSY models \cite{Kaplan} where de-couplings of the first and second 
generation sfermions are invoked to solve the SUSY FCNC and CP problems 
without spoiling the naturalness condition.
Another possibility is to arrange for partial cancellations among 
various contributions to the electron and neutron EDM's \cite{IN}.
Following the suggestions that the phases do not have to be suppressed,
many important works on the effects due to the phases have been already 
reported; the effects are very significant in extracting 
the parameters in the SUSY Lagrangian from experimental data \cite{SYCHOI}, 
estimating dark matter densities and scattering cross sections and 
Higgs boson limits \cite{BK,FO,PW}, CP violation in the $B$ and $K$ 
systems \cite{Ko}, and so on. 

The reaction $p\bar{p}\rightarrow\tilde{\chi}^\pm_1\tilde{\chi}^0_2$ is
one of the cleanest SUSY processes at the Tevatron. The subsequent
leptonic decays of charginos and neutralinos can lead to clean, {\it i.e.}
jet--free tri--lepton plus $\not\!\!\!{E}_T$ events which have very low
standard model backgrounds. Since charginos and neutralinos are expected
to be lighter than gluinos in models where gaugino masses unify near the
GUT scale, the clean tri--lepton signals potentially offer the largest
reach and the leptonic decays of $\tilde{\chi}^\pm_1$ and $\tilde{\chi}^0_2$
are not suppressed. These promising aspects of the SUSY tri--lepton signals
have led to a lot of detailed theoretical investigations \cite{MOST} 
and they have been confirmed by several simulation works \cite{Paige}. 
However, most of the works
have been done under the assumption that all the couplings are related at 
the grand unification or Planck scale and they are real. 
In this letter, in light of the possibility of large 
CP phases in the general MSSM, we re--visit the SUSY tri--lepton
signals at the Tevatron in the framework of MSSM assuming CP phases 
which are also severely constrained by EDM's.

The SUSY parameter set of the electroweak gaugino sector in the MSSM is 
$\{|M_1|, \Phi_1, M_2, |\mu|, \Phi_\mu, \tan\beta\}$. Note that
the phase of the SU(2) gaugino mass $M_2$ is rotated away by field
redefinitions so that $M_2$ is assumed real and positive without loss of 
any generality while the phases $\Phi_1$ and $\Phi_\mu$ of the U(1) 
gaugino mass $M_1$ and higgsino mass parameter $\mu$ remain as physical 
phases \cite{Note1}. For the sfermion sector, we assume the flavor--diagonal
sfermion mass matrices, a universal soft--breaking mass $m_{\tilde{f}_{L,R}}$
for sfermions of each chirality, and a universal trilinear term $A$. 
The universality condition enables us to have a controllable number of 
parameters while the effects due to its violation can be kept small by 
taking rather large sfermion masses and trilinear couplings in the general 
case with more complicated sfermion sectors. 

Since the CP phases might give a significant contribution to the electron and
neutron EDM's, it is incumbent to take into account the constraints on the 
CP phases of the EDM's, in particular, the electron EDM,
which are known to be stronger than those of the neutron EDM \cite{IN}. 
Related with the electron EDM constraints ($|d_e|\leq 4.3\times 10^{-27} 
e\cdot{\rm cm}$ at 95\% confidence level \cite{eEDM}), we investigate 
two distinct scenarios for the higgisno mass parameter 
$|\mu|$ and the slepton and squark masses:
\begin{eqnarray}
&& {\cal S}1: \, |\mu|= .2\, {\rm TeV}, 
                \, m_{\tilde{l}_{L,R}}= 10\, {\rm TeV}, 
                \, m_{\tilde{q}_{L,R}}= 10\, {\rm TeV}, \nonumber\\
&& {\cal S}2: \, |\mu|=.7\,  {\rm TeV},
                \, m_{\tilde{l}_{L,R}}= .2\, {\rm TeV}, 
                \, m_{\tilde{q}_{L,R}}= .5\, {\rm TeV}.
{ }
\label{eq:parameter}
\end{eqnarray}
In addition, we take a small value of $\tan\beta=3$ and $M_2=100$ GeV,
assume the gaugino mass unification $|M_1|=\frac{5}{3}\tan^2\theta_W M_2
\approx 0.5 M_2$ only for the magnitude, and take $A=1$ TeV for the
universal trilinear term while its phase is left arbitrary. 
Two CP phases $\Phi_\mu$ and $\Phi_1$ relevant for the tri--lepton
signal are strongly restricted in ${\cal S}2$ by the electron EDM \cite{IN} 
while they are not in ${\cal S}1$ with very heavy first and 
second--generation sfermions suggested by the effective SUSY models, 
unless $\tan\beta$ is very large and at the same time the third--generation 
sfermions are light \cite{Pilaftsis}. 
A larger value of $|\mu|=.7$ TeV in ${\cal S}2$ than in ${\cal S}1$ 
is chosen so that a relatively large space is allowed for the CP 
phases against the electron EDM constraints \cite{IN}. 
Squark masses are taken to be larger than slepton masses in 
${\cal S}2$, which is compatible with the sfermion spectrum in 
the minimal supergravity framework. The constraints imposed 
on the phases by the electron
EDM data are for some particular values of soft parameters with relatively 
light mass spectra in ${\cal S}2$. Nevertheless, we expect that
the results will exhibit general features typical for similar choices. 

Although the off--diagonal elements in the sfermion
mass matrices are proportional to small Yukawa couplings,
the trilinear terms are very crucial for the electron and neutron 
EDM's because the contributions of their phases to the EDM's require 
a chirality flip leading to dipole moments proportional to 
the relevant mass. Therefore, it is necessary to maintain the
sfermion left--right mixing in evaluating the EDM's. 
On the other hand, a relatively small $\tan\beta=3$ along with large 
sfermion masses tends to degrade the importance 
of sfermion left--right mixing effects except the effects from the stop sector
in the tri--lepton process. Without generational mixing, the parameters 
related to the third--generation 
particles are not directly involved in the tri--lepton process although they 
affect the decay branching fractions indirectly as well as
the electron EDM through two--loop diagrams \cite{Pilaftsis}.
Therefore, we can safely neglect the sfermion left--right mixing 
in evaluating tri--lepton event rates \cite{Note2} and concentrate 
mainly on the impact of two CP phases $\{\Phi_1$, $\Phi_\mu\}$, and 
the SU(2) gaugino and higgsino parameters $M_2$ and $|\mu|$ on the 
tri--lepton signals at the Tevatron with only the $e$ and $\mu$ as 
the final--state leptons \cite{Note3}. We emphasize in passing that 
the formidable hadronic backgrounds at the Tevatron experiments prevent
one from using the hadronic decays of charginos and neutralinos 
as useful search modes unlike at clean $e^+e^-$ collider experiments.

Figure~1 shows the mass spectrum of the lightest chargino 
$\tilde{\chi}^\pm_1$ and the neutralinos $\tilde{\chi}^0_{1,2}$ in 
the two scenarios; ${\cal S}1$ and ${\cal S}2$ in Eq.~(\ref{eq:parameter}). 
The upper three figures are for ${\cal S}1$ with large sfermion masses 
of 10 TeV and the lower three figures for ${\cal S}2$ where the shadowed areas
denote the region excluded by the electron EDM constraints. 
Except for the region of $\Phi_\mu=0,2\pi$ in ${\cal S}1$, 
$m_{\tilde{\chi}^\pm_1}$ and $m_{\tilde{\chi}^0_2}$ are very similar
in size and independent of $\Phi_1$ in both ${\cal S}1$ and ${\cal S}2$
while $m_{\tilde{\chi}^0_1}$ exhibits a very strongly correlated
dependence on the phases. The chargino mass $m_{\tilde{\chi}^\pm_1}$
increases as $\Phi_\mu$ approaches $\pi$, while $m_{\tilde{\chi}^0_1}$
becomes maximal at non--trivial values of  $\Phi_\mu$ and $\Phi_1$ in 
${\cal S}1$ with $|\mu|=200$ GeV. This implies that $m_{\tilde{\chi}^0_1}$ is 
strongly affected by a small value of $|\mu|$, while $m_{\tilde{\chi}^\pm_1}$ 
and $m_{\tilde{\chi}^0_2}$ are essentially determined by the SU(2)
gaugino mass $M_2$. Furthermore, a small $|\mu|$ tends to reduce 
$m_{\tilde{\chi}^0_{1,2}}$ and $m_{\tilde{\chi}^\pm_1}$ on the whole.

The decay patterns for the charginos and neutralinos are very much
parameter--dependent as well. The chargino decay $\tilde{\chi}^-_1
\rightarrow \tilde{\chi}^0_1\ell^-\bar{\nu}_l$ occurs though the 
$W$--exchange, slepton and sneutrino exchanges. For the sneutrino 
and slepton  much heavier than the chargino, the $W$--exchange 
contribution dominates, and the decay branching ratios among the 
leptonic modes are determined by those of the 
on--shell $W$ boson. Similarly, the 3--body decay $\tilde{\chi}^0_2
\rightarrow\tilde{\chi}^0_1\ell^+\ell^-$ occurs through virtual $Z$ bosons 
and sleptons. For the sleptons much heavier than the neutralino, 
the neutralino decays proceed through $Z^*$ with branching ratios similar 
to those of the on--shell $Z$ boson. We calculate the semileptonic 
branching fractions fully incorporating all the possible decay modes 
of the chargino and neutralino and find that the branching fractions are 
extremely sensitive to $\Phi_\mu$ and $\Phi_1$, especially in the scenario
${\cal S}2$ with a large value of $|\mu|$. 
Figure~2 shows ${\cal B}(\tilde{\chi}^-_1\rightarrow\tilde{\chi}^0_1
\ell^-\nu_l)$ and ${\cal B}(\tilde{\chi}^0_2\rightarrow\tilde{\chi}^0_1
\ell^+\ell^-)$ for $\ell=e$ or $\mu$ in ${\cal S}1$ (two upper figures) 
and in ${\cal S}2$ (two lower figures). 
${\cal B}(\tilde{\chi}^-_1\rightarrow\tilde{\chi}^0_1\ell^-\nu_l)$ is 
almost constant over the whole space of the phases in ${\cal S}1$,
while ${\cal B}(\tilde{\chi}^0_2\rightarrow\tilde{\chi}^0_1\ell^+\ell^-)$ 
is very sensitive to $\Phi_1$ around $\Phi_\mu=0,2\pi$. 
On the other hand, both branching fractions strongly depend on $\Phi_\mu$ 
and $\Phi_1$ in ${\cal S}2$. Remarkably ${\cal B}(\tilde{\chi}^-_1
\rightarrow\tilde{\chi}^0_1\ell^-\nu_l)$ is minimal for non--trivial
phases. ${\cal B}(\tilde{\chi}^0_2\rightarrow\tilde{\chi}^0_1\ell^+\ell^-)$ 
is enhanced in ${\cal S}2$ because the slepton-exchange contributions 
due to mainly the gaugino components of the neutralinos become dominant 
due to the small slepton masses and the large 
value of $|\mu|$. On the other hand, ${\cal B}(\tilde{\chi}^-_1\rightarrow
\tilde{\chi}^0_1\ell^-\nu_l)$ does not change so much in size, but the 
dependence of the branching fractions on the phases becomes very 
different in ${\cal S}2$.

The parton--level production process $d\bar{u}\rightarrow\tilde{\chi}^-_1
\tilde{\chi}^0_2$ is generated by the $s$--channel $W^-$ exchange, 
$t$--channel $\tilde{d}$ exchange and $u$--channel $\tilde{u}^*$ exchange.
We need to convolute an effective parton distribution with the
cross section of the parton--level process to obtain the total production 
cross section in $p\bar{p}$ collisions, $\sigma(p\bar{p}\rightarrow
\tilde{\chi}^-_1\tilde{\chi}^0_2+{\rm X})$. For our analysis we use
the CTEQ4m parton distribution function \cite{CTEQ4m} with the QCD 
scale of the c.m. energy of the parton--level process and with
the dominant QCD radiative corrections included by taking the value of 
the enhancement factor $\kappa=1.3$ \cite{SPIRA}. 
The production cross section $\sigma(p\bar{p}\rightarrow
\tilde{\chi}^+_1\tilde{\chi}^0_2+{\rm X})$
for the positive chargino and neutralino pair
in the CP self-conjugate $p\bar{p}$ collisions is the same as its 
charge--conjugate one. The two upper figures in Fig.~3 shows the dependence 
of $\sigma(p\bar{p}\rightarrow\tilde{\chi}^-_1\tilde{\chi}^0_2+{\rm X})$ on 
$\Phi_\mu$ and $\Phi_1$ in (a) ${\cal S}1$ and (b) ${\cal S}2$. 
Note that except for the region of $\Phi_\mu=0,2\pi$ in ${\cal S}1$, the
cross section is almost independent of the phase $\Phi_1$. The large
$|\mu|$ and small squark masses in ${\cal S}2$ reduce the production cross
section due to a destructive interference between the $W$-exchange
and the squark--exchange diagrams. We find that in both cases
$\sigma(p\bar{p}\rightarrow\tilde{\chi}^-_1\tilde{\chi}^0_2+{\rm X})$ 
decreases as $\Phi_\mu$ approaches $\pi$.

A realistic analysis for the tri--lepton signal demands a numerical 
simulation fully incorporating all the background processes, for which
one needs to make a quite considerable investigation. We defer 
such a detailed analysis to our next work \cite{NEXT} and present the 
total event rate of the tri--lepton signal without any experimental 
cuts against possible backgrounds. The lower two figures in Fig.~3 show 
the dependence of the total cross section $\sigma(p\bar{p}\rightarrow 
3\ell+{\rm X})$ on $\Phi_\mu$ and $\Phi_1$ 
in (a) ${\cal S}1$ and (b) ${\cal S}2$, which can be obtained by 
multiplying $\sigma(p\bar{p}\rightarrow\tilde{\chi}^-_1\tilde{\chi}^0_2)$ 
with ${\cal B}(\tilde{\chi}^-_1\rightarrow\tilde{\chi}^0_1\ell^-\nu)$
and ${\cal B}(\tilde{\chi}^0_2\rightarrow\tilde{\chi}^0_1\ell^+\ell^-)$. 
Since ${\cal B}(\tilde{\chi}^-_1\rightarrow\tilde{\chi}^0_1\ell^-\nu)$
remains almost constant in ${\cal S}1$ as shown in Fig.~2, the total 
cross section is mainly affected by ${\cal B}(\tilde{\chi}^0_2\rightarrow
\tilde{\chi}^0_1\ell^+\ell^-)$ and so it strongly depends on $\Phi_1$ 
around $\Phi_\mu=0,2\pi$. Note that the total tri--lepton cross section 
is ${\cal O}(10\,{\rm fb})$, too small to be seen with the present
accumulated luminosity at the Tevatron of ${\cal O}(0.1\, {\rm fb}^{-1})$. 
However, the future Tevatron experiments with its upgraded luminosity of 
$2\,{\rm fb}^{-1}$ may exclude the region around $\Phi_\mu=0,2\pi$ 
and $\Phi_1=\pi$, if SUSY is not discovered. In ${\cal S}2$, the total 
cross section depends very strongly on the CP phases, takes its minimum 
value for  non--trivial 
CP phases, and is larger than that in ${\cal S}1$ due to the largely 
enhanced neutralino branching fraction as shown in Fig.~2 which surpasses
the reduction due to the destructive interference in the production 
cross section. Depending on the integrated luminosity, therefore, 
the very existence of the minimum event rate and the simultaneous small 
mass splitting for non--trivial phases as can be checked in Fig.~1 
reflect that the range of the chargino and neutralino masses, 
which could be ruled out at the Tevatron, might be {\it much} smaller 
than that \cite{CCEFM} ruled out in the context of SUGRA and GUT inspired 
SUSY models. 

To summarize, we have investigated the impact of the phases $\Phi_\mu$ 
and $\Phi_1$ on the SUSY tri--lepton signals at the Tevatron in the MSSM
with general CP phases without generational mixing under the constraints
on the phases by the electron EDM data. For the sake of generality, we have
considered two exemplary scenarios for the relevant SUSY parameters; 
${\cal S}1$ with very heavy first-- and second--generation sfermions
and ${\cal S}2$ with relatively light sfermions but a large $|\mu|$.
We have found that in both scenarios the CP phases significantly  affect 
the production cross section and especially 
the partial leptonic branching fractions of the chargino $\tilde{\chi}^\pm_1$
and neutralino $\tilde{\chi}^0_2$. As a result, there may lead to a 
{\it minimum rate of the tri--lepton signal for non--trivial CP phases}. 
This implies that one should be careful when interpreting the chargino and
neutralino mass limits derived under the assumption of vanishing phases, 
since the worst case is not (always) covered by just flipping 
the sign of $\mu$; rather it can occur from some non--trivial phases in 
between. 

\vskip 0.3cm


We are grateful to Manuel Drees and Peter Zerwas for valuable comments and
helpful discussions.
This work was supported in part by the Korea Science and Engineering 
Foundation (KOSEF) through the KOSEF--DFG large collaboration project, 
Project No.~96--0702--01-01-2, and in part by the Center for 
Theoretical Physics. MG acknowledges Alexander von Humboldt Stiftung 
foundation for financial help and also KOSEF for funding during his stay 
in Yonsei University, Seoul, where this work was initiated.

\vskip 1.7cm

\begin{figure}
 \begin{center}
\hbox to\textwidth{\hss\epsfig{file=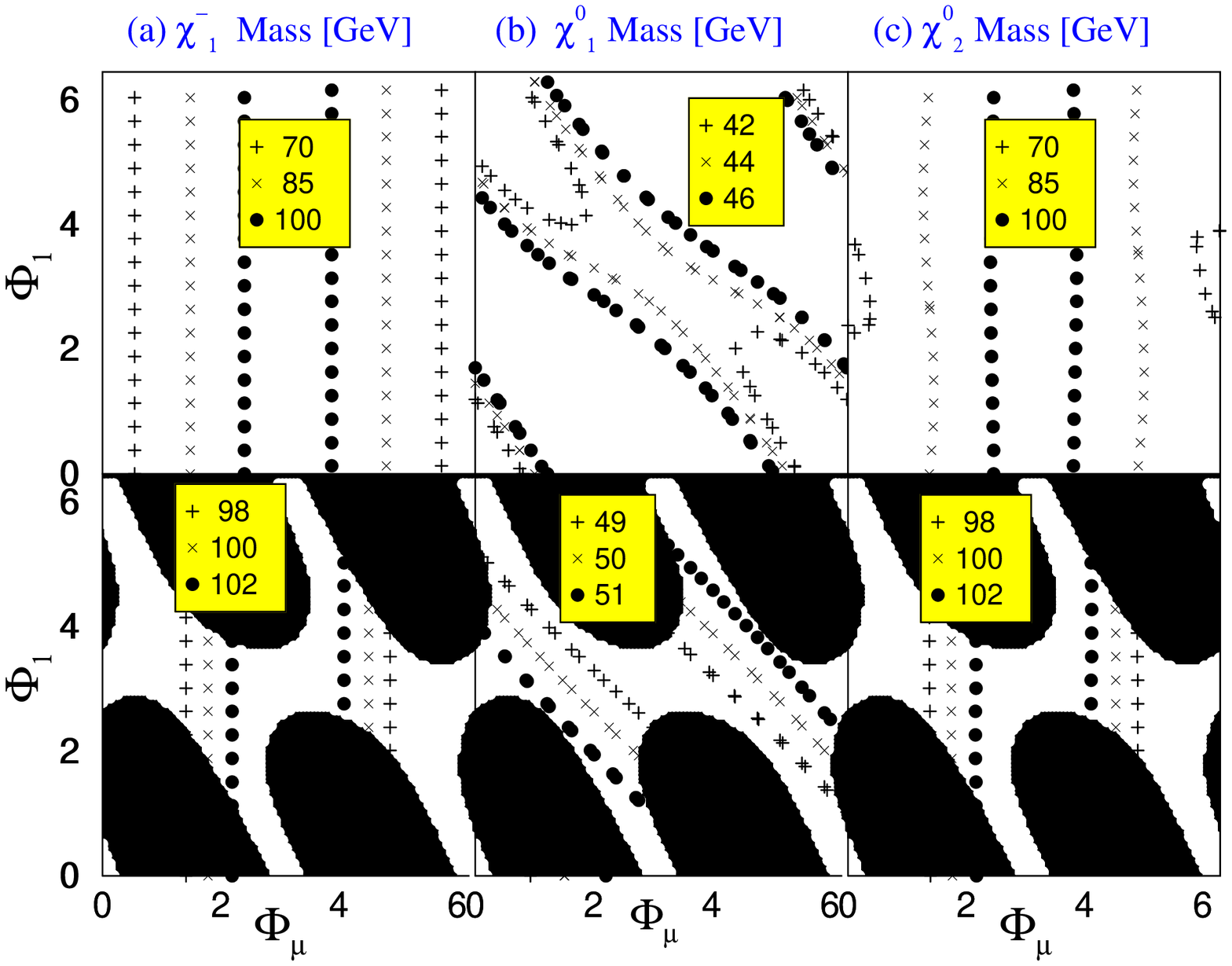,width=16cm,height=14cm}\hss}
 \end{center}
\caption{(a) $m_{\tilde{\chi}^-_1}$, (b) $m_{\tilde{\chi}^0_1}$ and
         (c) $m_{\tilde{\chi}^0_2}$ on the $\{\Phi_\mu,\Phi_1\}$ plane
              in ${\cal S}1$ (upper part) and ${\cal S}2$ (lower part).}
\label{fig1}
\end{figure}

\vskip 1.5cm

\begin{figure}
 \begin{center}
\hbox to\textwidth{\hss\epsfig{file=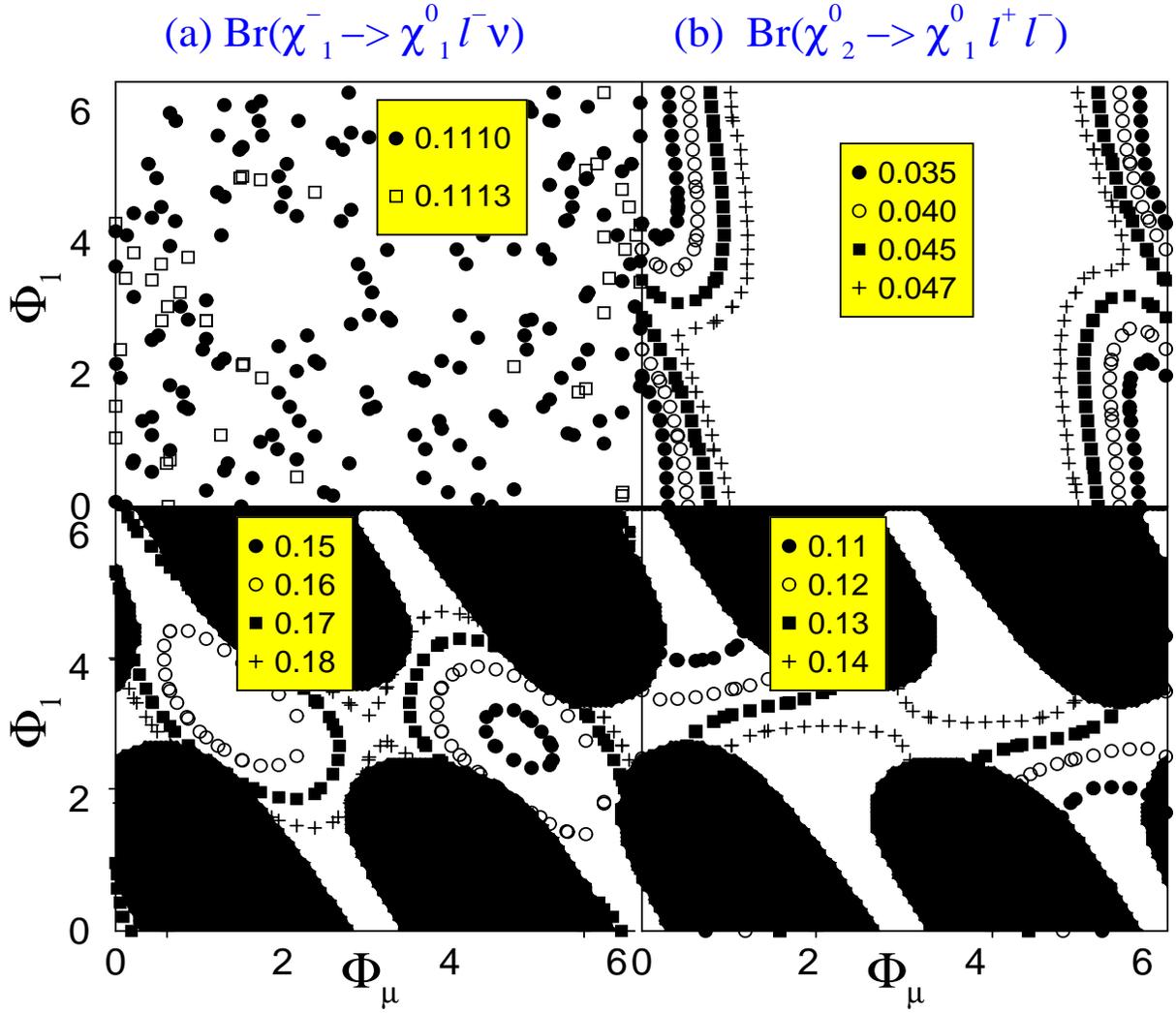,width=16cm,height=14cm}\hss}
 \end{center}
\caption{${\cal B}(\tilde{\chi}^-_1\rightarrow\tilde{\chi}^0_1\ell^-\nu_l)$ 
         and ${\cal B}(\tilde{\chi}^0_2\rightarrow\tilde{\chi}^0_1
         \ell^+\ell^-)$ for $l=e$ or $\mu$ on the $\{\Phi_\mu,\Phi_1\}$ plane 
         in the scenarios ${\cal S}1$ (upper part) and ${\cal S}2$ 
         (lower part).}
\label{fig2}
\end{figure}

\vskip 1cm

\begin{figure}
 \begin{center}
\hbox to\textwidth{\hss\epsfig{file=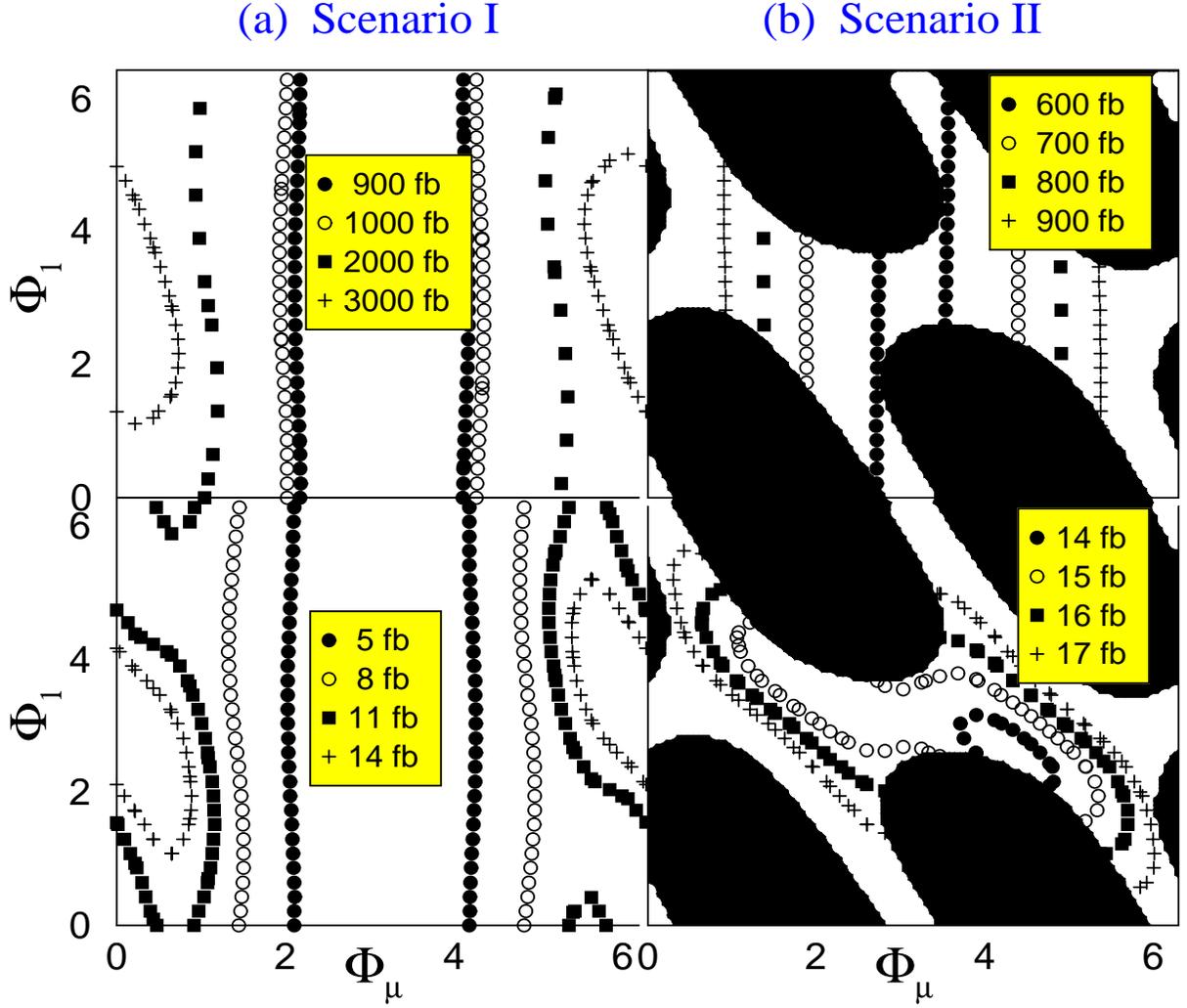,width=16cm,height=14cm}\hss}
 \end{center}
\caption{$\sigma(p\bar{p}\rightarrow\tilde{\chi}^-_1\tilde{\chi}^0_2+
         {\rm X})$ (upper
         part) and $\sigma(p\bar{p}\rightarrow 3\ell + {\rm X})$ (lower part)
         on the $\{\Phi_\mu,\Phi_1\}$ plane in (a) ${\cal S}1$ and (b) 
         ${\cal S} 2$.}
\label{fig3}
\end{figure}

\vfil\eject

\end{document}